\documentclass[10pt,a4paper]{article}

\usepackage{amsmath,amssymb,amscd,latexsym}
\usepackage{stmaryrd}
\usepackage{graphics, psfig,graphicx}
\usepackage{eufrak}
\usepackage{multind}

\usepackage[all,dvips]{xy}

\usepackage{theorem}

\numberwithin{equation}{section}
\newcommand\refeq[1]{(\ref{#1})}

\newtheorem{def.notation}{D\'efinition--Notations}[section]
\newtheorem{defff}{Definition}[section]
\newtheorem{theo}{Th\'eor\`eme}[section]

\newtheorem{prop}{Proposition}[section]
\newtheorem{prop.def}{Proposition--D\'efinition}[section]

\newtheorem{propriete.def}{Propri\'et\'e--D\'efinition}[section]

\newtheorem{lemme}{Lemme}[section]

\newenvironment{proof}{\noindent{\textsf{\underline{Proof}: }}}{$\blacksquare$}

\newtheorem{theorem}{Theorem}[section]
\newtheorem{theorem*}{Theorem}

\theoremstyle{break}
{\theorembodyfont{\rmfamily}}
{\theorembodyfont{\rmfamily}}
{\theorembodyfont{\rmfamily}}
{\theorembodyfont{\rmfamily}}
{\theorembodyfont{\rmfamily}}
{\theorembodyfont{\rmfamily}}
{\theorembodyfont{\rmfamily}}
{\theorembodyfont{\rmfamily}\newtheorem{remark}{Remark}[section]}
{\theorembodyfont{\rmfamily}}

\theoremstyle{plain}
\newtheorem{remark.num}{\sous{Remark}: }[section]

\newcommand\ie{i.e. }
\newcommand\eg{e.g. }

\newcommand\trema[1]{\"{#1}}

\newcommand\N{\mathbb{N}}
\newcommand\R{\mathbb{R}}

\newcommand\C{\mathbb{C}}
\newcommand\perm{\mathfrak{S}}

\newcommand\D{\partial}

\newcommand\dd{\text{d}}
\newcommand{\inte}{\hbox{ {\leaders\hrule\hskip.2cm}{\vrule height .22cm} }}

\newcommand\opa{\textbf{\large{a}}}

\newcommand\lent{\llbracket }
\newcommand\rent{\rrbracket }

\newcommand\vect[1]{\overrightarrow{#1}}
\newcommand\sous[1]{\underline{#1}}
\newcommand\sur[1]{\overline{#1}}
\newcommand\dsurd[2]{\frac{\partial #1}{\partial #2} }
\newcommand\chapo[1]{\widehat{#1} }
\newcommand\mc[1]{\ensuremath{\mathcal{#1}} }

\newcommand\sym[1]{S(#1)}
\newcommand\Min{\mc{X}_{0}}

\newcommand\PP{P} 

\newcommand\Fock{\mc{F}_s}

\title{Covariant Deformation Quantization of Free Fields}
\author{Dikanaina HARRIVEL\footnote{LAREMA, UMR 6093, Universit\'e d'Angers, France.  
\textbf{dika@tonton.univ-angers.fr}}}

\begin{document}

\maketitle

\begin{abstract}
We define covariantly a deformation of a given algebra, then we will see how it can be related to a deformation quantization of 
a class of observables in Quantum Field Theory. Then we will investigate the operator order related to this deformation quantization. 
\end{abstract}

\vspace{ 1cm} 

\section{introduction}
One of the purpose of multisymplectic geometry is to give a Hamiltonian formulation of (classical) 
field theory similar to the symplectic formulation of the one dimensional Hamiltonian formalism 
(the Hamilton's formulation of Mechanics). The time variable is replaced by space-time variables and the cotangent bundle by 
a finite dimensional manifold equipped with a \emph{multisymplectic form} similar to the canonical symplectic form on the 
cotangent bundle. Starting from a Lagrangian density which describes the dynamics of the field, one can construct a Hamiltonian 
function through a Legendre transform and obtain a geometric formulation of the problem. 
For an introduction to the multisymplectic geometry one can refer to \cite{sujet} and for more complete 
informations one can read the papers of F. H\'elein and J. Kouneiher \cite{HK1}, \cite{HK2}.

Note that this formalism differs from the standard Hamiltonian formulation of field theory used by physicists 
(see \eg \cite{ITZAK}), in particular the multisymplectic approach is covariant \ie compatible with the principles of 
special and general Relativity and everything is finite dimensional. 

Then we have to define the observable quantities and the Poisson bracket between these observables quantities. 
A notion of observable has been defined by the polish school in the seventies \cite{Tul.observables}, \cite{Kij}, \cite{HK1}, 
\cite{HK2}. The set of these quantities becomes a Poisson algebra just as in the one dimensional case. The next task is to quantize 
it. The purpose of this paper is to propose a quantization using the deformation quantization procedure. 

First we will present a deformation of a given algebra ; then we will show how it is connected 
to Field Theory. Then we will see how to recover this deformation quantization by means of operator ordering in the quantization 
procedure. We will see that this ordering differs from the usual Wick order (see \eg \cite{ITZAK}) used by physicist.  Finally 
we will see how the deformation quantization introduced by J. Dito \cite{Dito}, \cite{Dito2} (which corresponds to the Wick 
order) can be applied to our problem. 

\section{Deformation}
Let $M$ denote the space $M:=\R^{n+1}$ endowed with a metric $g^{\beta\alpha}$ and $\omega$ denote a volume form on $M$. 
We will denote by $(x^\beta)_{\beta\in\lent 0,n\rent}$ a system of coordinates on $M$ such that 
$\omega=\dd x^0\wedge\cdots\wedge\dd x^n$ and by $V$ the real vector space defined by 
$$
V:=\left\{
\psi\in\mc{C}^\infty (M)\ ;\ g^{\beta\alpha}\frac{\D^{2}\psi}{\D x^\alpha\D x^\beta}+m^2\psi=0
\right\}
$$
Then one can construct $(SV,\odot)$ the free commutative algebra generated by $V$ (see \eg \cite{alg.symetrique}) where $\odot$ 
denotes the associative and commutative symmetric tensor product. 
Then $SV$ have a natural structure of a graded algebra $SV:=\bigoplus_{p\ge 0}S^pV$ where $S^0V=\R$ and for $p\ge 1$, 
$S^pV$ denotes the $p$--th symmetric tensor of $V$. 

Consider the \emph{Spencer operator} $\delta:SV\to V\otimes SV$ \ie the linear operator such that $\delta (S^0 V)=0$ and 
for any decomposable element $v=v_1\odot \cdots \odot v_k$, 
$\delta v:=\sum_i v_i\otimes v_1\odot \cdots\odot v_{i-1}\odot v_{i+1}\odot \cdots \odot v_k$. Then we introduce the following 
definition 
\begin{defff}\label{def.i.d}\sl{
\begin{itemize}
\item
Let $\beta$ be in $\lent 0,n\rent$ and $\eta\in M$. We denote by $i(\eta)$ and $\D_\beta(\eta)$ the linear operators acting on $SV$ 
defined by  
$$
\begin{array}{lcr}
\displaystyle{i(\eta):=\left(\phi(\eta)\otimes id\right)\circ \delta}&\text{ and }&
\displaystyle{\D_\beta(\eta):=
\left(\dsurd{\phi(\eta)}{x^\beta}\otimes id \right)\circ \delta}
\end{array}
$$
where $\phi(\eta)$ and $\dsurd{\phi(\eta)}{x^\beta}$ denote the real valued operators on $V$ such that 
$\phi(\eta)(\psi):=\psi(\eta)$ and $\dsurd{\phi(\eta)}{x^\beta}(\psi):=\dsurd{\psi}{x^\beta}(\eta)$. 
\item 
Let $p\in\N^*$, $\beta=(\beta_1,\ldots,\beta_p)\in\lent 0,n\rent^p$ and $\eta=(\eta_1,\ldots,\eta_p)\in M^p$. 
We define the operators $i^p(\eta)$ and $\D^p_{\beta}(\eta)$ acting on $SV$ by 
$$
\left\{
\begin{array}{l}
\displaystyle{i^p(\eta):=i(\eta_p)\circ \cdots\circ i(\eta_1)}\\
\displaystyle{\D^p_{\beta}(\eta):=\D_{\beta_p}(\eta_p)\circ \cdots\circ\D_{\beta_1}(\eta_1)}
\end{array}
\right.
$$
and we impose $i^0=\D^0=Id_V$.
\end{itemize}
}
\end{defff}
Let $SV[[\hbar]]$ be the space of formal power series with coefficients in $SV$, and let us fix some hypersurface $\Sigma$ on 
$M$. Then we define a star product $\star_\hbar $ on $SV[[\hbar]]$ by the following 
\begin{defff}\label{def.star}\sl{
We denote by $\star_\hbar$ the $\R[[\hbar]]$--module morphism $\star_\hbar :SV[[\hbar]]\otimes SV[[\hbar]]\to SV[[\hbar]]$ such 
that for all $\Psi\in SV$ and $\Phi\in SV$, 
$\Psi\star_\hbar \Phi:=\sum_{p\ge 0}\hbar ^p\frac{1}{p!}\left[ \Psi \star_\hbar \Phi\right]_p$ 
where $\left[\Psi \star_\hbar\Phi\right]_0:=\Psi\odot\Phi$ and where for all $p\in\N^*$
$$
\left[\Psi \star_\hbar\Phi\right]_p:=\sum_{\substack{\beta\in\lent 0,n\rent^p\\ \alpha\in\lent 0,n\rent^p}}
g^{\alpha\beta}\int_{\Sigma^p}\left(\D^p_{\beta}(\eta)\Psi\right)\odot\left(i^p(\eta)\Phi\right)
\omega_{\alpha}(\eta)
$$
Here for all $\alpha=(\alpha_1,\ldots,\alpha_p)$ and $\beta=(\beta_1,\ldots,\beta_p)$ in $\lent 0,n\rent^p$, $g^{\beta\alpha}$ 
denotes $g^{\beta_1\alpha_1}\cdots g^{\beta_p\alpha_p}$ and 
$\omega_{\alpha}:=\omega_{\alpha_1}\otimes\cdots\otimes\omega_{\alpha_p}$ and for all $\mu\in\lent 1,n\rent$, we have denoted by 
$\omega_\mu$ the $n$--form $\dsurd{}{x^\mu}\inte\omega$. 
}
\end{defff}

\begin{prop}\label{deformation}\sl{
$(SV[[\hbar]],\star_\hbar )$ is an associative algebra, and this algebra is a \emph{deformation} of the algebra $(SV,\odot)$, 
\ie the projection $\pi:SV[[\hbar]]\longrightarrow SV$ is an algebra morphism.
}
\end{prop}

\begin{remark}
Note that since $\D^p_{\beta}(\eta)(S^k V)\subset S^{k-p}V$ and $\D^p_{\beta}(\eta)(S^k V)\subset S^{k-p}V$ 
(where by convention $S^lV=0$ for all $l<0$), the space $SV[\hbar]$ of polynomials with coefficients in $SV$ is a subalgebra 
of $(SV[[\hbar]],\star_\hbar )$
\end{remark}

We postpone the proof of proposition \ref{deformation} until the appendix \ref{dem.1}. 
Now we will see how the star prodduct $\star_\hbar$ can be related to the operator product of Quantum Field Theory. 

\section{Free Scalar Fields}
We denote by $\mc{F}(V,\R)$ the real valued functional on $V$. 
Let us consider the algebra morphisme $I:SV\longrightarrow\mc{F}(V,\R)$ such that for all $\psi\in V$ 
$$
I(\psi):\varphi\in V\longrightarrow\int_\Sigma g^{\alpha\beta}\left(\dsurd{\varphi}{x^\alpha}\psi-\varphi
\dsurd{\psi}{x^\alpha}\right)
$$
and denote by $\mc{L}$ the image of $I$ \ie $\mc{L}:=I(SV)$. The algebra $\mc{L}$ is the observable functionals of the covariant 
Hamiltonian formalism described by F. H\'elein and J. Kouneiher (see \cite{sujet} or \cite{HK1} for more informations on this 
formalism and multisymplectic geometry).   

For all $(\psi,\chi)\in V^2$ we define the brackets $\{\psi,\chi\}$ between $\psi$ and $\varphi$ by 
$$
\{\psi,\chi\}:=I(\psi)(\varphi)\in\R=S^0V
$$
Once again this bracket comes from multisymplectic geometry and it coincides with the brackets used by physicist in field theory 
(see \cite{HK1}, \cite{ITZAK}, \cite{RYDER}). 
These brackets admit unique extension to $SV$ satisfying Leibnitz rule \ie such that 
$\{a\odot b,c\}=a\odot\{b,c\}+\{a,c\}\odot b$ for all $(a,b,c)\in SV^3$ (see \eg \cite{Brouder.Oeckl}). 

\begin{prop}\label{correspondance}\sl{
For all $(\psi,\chi)\in SV^2$ we have $\pi\left(\frac{1}{\hbar}[\psi,\chi]_{\hbar}\right)=\{\psi,\chi\}$ where 
$[\cdot,\cdot]_{\hbar}$ denotes the commutator on $SV[[\hbar]]$. Hence $(SV,\odot,\{\bullet,\bullet\})$ is a Poisson algebra 
\ie $(SV,\odot)$ is an algebra and the bracket satisfies Leibnitz rule and Jacobi identity : 
$\{a,\{b,c\}\}+\{c,\{a,b\}\}+\{b,\{c,a\}\}=0$. Moreover $(SV[[\hbar]],\star_\hbar)$ is a deformation quantization of 
$(SV,\odot,\{\bullet,\bullet\})$. 
}
\end{prop}

Hence we have defined a deformation quantization of the Poisson algebra of observables quantities which arise in multisymplectic 
geometry. This deformation quantization can be generalized to more general space--time manifold and we think that it can have some 
multisymplectic interpretation. 

Unfortunately when we try to link our star product with canonical free quantum field theory we see that it correspond with 
an operator ordering which differs from the \emph{Wick ordering} or \emph{normal ordering} used by physicists 
(see \cite{ITZAK}, \cite{RYDER}, \cite{Fock.rigoureux.1} or \cite{Fock.rigoureux.2}). 

%

\subsection{Operator ordering}\label{section.ordre}
Hence suppose that the metric $g^{\beta\alpha}$ is given by the diagonal matrix $(-1,1,\ldots,1)$ with respect to the coordinates 
$(x^\beta)_\beta$. The first variable plays the role of time and we will denote it by $t$ and we consider the hypersurface 
$\Sigma$ of $M$ defined by $\Sigma=\{x\in M\ ;\ t=0\}$. In this case one can easily show that for all decomposable 
element $\Psi^{(k)}=\psi_1\odot\cdots\odot \psi_k\in S^kV$ of $SV$, the functional $I(\Psi^{(k)})$ is given by 
$\forall \varphi\in V$ 
\begin{equation}\label{I}
I(\Psi^{(k)})(\varphi)=\sum_{I\subset\lent 1,k\rent}(-1)^{\vert I\vert}
\left(\prod_{i\in I}\int_{t=0}\psi_i\dsurd{\varphi}{t}\right)\left(\prod_{j\not\in I}\int_{t=0}\dsurd{\psi_j}{t}\varphi\right)
\end{equation}
When we quantize this functional, we replace all the integral $\int_{t=0} \dsurd{\psi_j}{t}\varphi$ by 
an operator $\varphi_m\left(\dsurd{\psi_j}{t}(0,\bullet)\right)$ acting on a Hilbert space (\emph{Fock space}) and 
$\int_{t=0} \psi_j\dsurd{\varphi}{t}$ by an other operator $\pi_m(\psi_j(0,\bullet))$ acting on the same hilbert. For 
all $(f,g)\in L^2(\R^n)^2$ the operators $\varphi(f)$ are supposed to satisfy the \emph{canonical commutation relation} (CCR) : 
$\left[\varphi_m(f),\varphi_m(g)\right]=\left[\pi_m(f),\pi_m(g)\right]=0$ and  
\begin{equation}\label{CCR}
\left[\varphi_m(f),\pi_m(g)\right]=i\left(\int_{\R^{n-1}} fg\right)id
\end{equation}
One can refer to the book of Michael Reed and Barry Simon \cite{Fock.rigoureux.2} p.210--218 for a definition of $\varphi_m(f)$ 
and $\pi_m(g)$. We denote by $\mc{O}$ the set of operators acting on $\mc{F}_s$. 

But since the operators $\varphi_m(f)$ and $\pi_m(g)$ don't commute, there is not a unique way of replacement. This is the 
problem of operator ordering. Following J. Dito \cite{Dito2},\cite{Dito} or K. Fredenhagen and M. D\"utsch 
\cite{Fredenhagen.Dutsch} to each deformation quantization corresponds an operator ordering. 

Let us choose to put all the operators $\pi_m(g)$ on the left of the operators $\varphi_m(f)$ \ie consider the 
linear map $\Theta:SP\longrightarrow \mc{O}$ such that for all decomposable element $\Psi^{(k)}=\psi_1\odot \cdots\odot \psi_k$ 
of $S^k V$ 
$$
\Theta(\Psi^{(k)}):=\sum_{I\subset\lent 1,k\rent}(-1)^{\vert I\vert}
\prod_{i\in I}\pi_m\left(\psi_i(0,\bullet)\right)\prod_{j\not\in I}\pi_m\left(\dsurd{\psi_j}{t}(0,\bullet)\right)
$$
Then we define $\Theta$ on $SV[\hbar]$ by $\R[\hbar]$--linearity by setting $\Theta(\hbar):=-i$. Then we have the 
following result 
\begin{theo}\label{ordre}\sl{ 
Let $A$ and $B$ belong to $SV[\hbar ]$, then $\Theta(A)\Theta(B)=\Theta(A\star_\hbar B)$
}
\end{theo}
We postpone the proof of theorem \ref{ordre} until the appendix \ref{dem.2}. 
The theorem \ref{ordre} ensures that the operator ordering which we consider corresponds to our star product $\star_\hbar$. 
We see that this ordering differs from the usual normal order used in Free Quantum Field Theory. 
In the next section we define a star product (described by J. Dito in \cite{Dito} and \cite{Dito2}) which corresponds 
to the normal order. 

\subsection{Wick Order}
One can remark that the operator ordering corresponding to our product is not the usual order of Quantum Field Theory 
\cite{RYDER}, \cite{ITZAK}. We will see here the deformation quantization which corresponds to the normal order.  

For all decomposable element $\Psi^{(k)}=\psi_1\odot \cdots\odot \psi_k$ of $S^k V$ we denote $\Theta_W(\Psi^{(k)})$ the 
operator obtained by replacing the integral $\int_{t=0} \dsurd{\psi_j}{t}\varphi$ by 
$\varphi_m\left(\dsurd{\psi_j}{t}(0,\bullet)\right)$ and $\int_{t=0} \psi_j\dsurd{\varphi}{t}$ by 
$\pi_m(\psi_j(0,\bullet))$ in \refeq{I} and using the normal order product. Then one can show that  
\begin{equation}\label{Theta_W}
\Theta_W(\Psi^{(k)})=\sum_{I\subset\lent 1,k\rent} \left[\prod_{i\in I}\opa^{\dagger}(G\psi_i)\right]
\left[\prod_{j\not\in I}\opa(F\psi_j)\right]
\end{equation}
where for all $\psi\in\PP$, $F\psi$ and $G\psi$ denotes the functions 
\begin{equation*}
F\psi:=\frac{1}{\sqrt{2}}\left(1/\mu\chapo{\dsurd{\psi}{t}(0,\bullet)}-i\chapo{\psi(0,\bullet)}\right)\text{ and } 
G\psi:=\frac{1}{\sqrt{2}}\left(1/\mu\chapo{\dsurd{\psi}{t}(0,\bullet)}+i\chapo{\psi(0,\bullet)}\right)
\end{equation*}
Then we define $\Theta_W$ on $SP[\hbar]$ by $\R[\hbar]$--linearity by setting $\Theta_W(\hbar):=1$. 

Now let us define another star product $\star_W$ on $SV[\hbar]$ such that $\Theta_W$ is an algebra morphism between 
$(SV[\hbar],\star_W)$ and $(\mc{W},\cdot)$. 

\paragraph{Normal Product}
\begin{defff}\label{d-.i+}\sl{
\begin{itemize}
\item 
Let $\vect{k}\in\R^{n-1}$ then we denote 
$\mc{G}(k):SP[[\hbar]]\otimes\C\longrightarrow SP[[\hbar]]\otimes\C$ and 
$\mc{G}(k)^*:SP[[\hbar]]\otimes\C\longrightarrow SP[[\hbar]]\otimes\C$ the  $\C[[\hbar]]$--linear operators defined by 
$$
\begin{array}{lcr}
\displaystyle{\mc{G}(\vect{k}):=\left(\sqrt{\mu}G(\vect{k})\otimes id\right)\circ \delta}&\text{ et }&
\displaystyle{
\mc{G}(\vect{k})^*:=
\left(\sqrt{\mu}\sur{G(\vect{k})}\otimes id\right)\circ \delta}
\end{array}
$$
where for all ${k}\in\R^{n-1}$ the operator $\sqrt{\mu}G(\vect{k}):\PP\longrightarrow\C$ is given by 
$$
\sqrt{\mu}{G(\vect{k})}(\psi):=\frac{1}{\sqrt{2}}{\left(1/\mu\chapo{\dsurd{\psi}{t}}+i\chapo{\psi}
\right)}(\vect{k})
$$
\item Let $p\in\N$ and $k^{(p)}\in M^p$, we define $\mc{G}^p(k^{(p)})$ and $\mc{G}^p(k^{(p)})^*$ by 
$\mc{G}^0=\mc{G}^{0*}=Id$ and for all $p\ge 1$ 
$$
\left\{
\begin{array}{l}
\mc{G}^p(\vect{k}):=\mc{G}(\vect{k}_p)\circ\cdots\circ\mc{G}(\vect{k}_1)\\
\mc{G}^p(\vect{k})^*:=\mc{G}(\vect{k}_p)^*\circ\cdots\circ \mc{G}(\vect{k}_1)^*
\end{array}
\right.
$$
\end{itemize}
}
\end{defff}
One can easily show that the operators $\mc{G}(k)$ and $\mc{G}(k)^*$ satisfy to the same properties than $\D_\alpha^p(\eta)$ 
and $i^p(\eta)$. Hence we can easily adapt the proof of proposition \ref{deformation} to show the following result 
\begin{prop}\sl{
Let $\star_W$ denotes a bilinear operator acting on $SV[\hbar]\otimes\C$ such that for all $(P,Q)\in SV[\hbar]\otimes\C^2$ 
$$
P\star_W Q:=\sum_k\frac{1}{p!}\int_{(\R^{n-1})^p}\dd{k}\mc{G}^p(\vect{k})P\odot \mc{G}^p(\vect{k})^*Q
$$
Then $(SV[\hbar]\otimes\C,\star_W)$ is an associative algebra with unit, moreover it is a deformation quantization of 
$(SV\otimes\C,\odot,\{\cdot,\cdot\})$. 
}
\end{prop}
Then using the Wick theorem and following the way we prove the theorem \ref{ordre}, one can prove the following result
\begin{theorem}\label{quantification}\sl{
The operator $\Theta_W$ is an algebra morphism from $(SV[\hbar]\otimes\C,\star_W)$ to $\mc{O}$, \ie  
$\Theta_W(1)=1$ and for all $P,Q$ belonging to $SV[\hbar]\otimes\C$
$$
\Theta_W(P)\Theta_W(Q)=\Theta_W(P\star_W Q)
$$
}
\end{theorem}
Hence the star product $\star_W$ corresponds to the Wick ordering. One can show that $\star_W$ corresponds to the product 
defined by J. Dito in \cite{Dito}, \cite{Dito2}. 

\section*{Acknowledgements} 
This paper is a part of author PhD Thesis submitted at Angers university, UMR 6093 CNRS. 
The author is grateful to his advisors Fr\'ed\'eric H\'elein and Volodya Roubtsov for encouragement anf helpful discussions. 
He also thanks Joseph Kouneiher for careful reading of the manuscript and useful remarks. 

\appendix

\section{Free Quantum Field}\label{appendice.champ.libre}
Here we recall the Quantum Field Theory which corresponds to massive free fields. One can see \cite{Fock.rigoureux.1} 
or \cite{Fock.rigoureux.2} for more details. 

\subsection{Fock space}\label{espace.fock}
Let $p$ belong to $\N^*$, then we denote by $L^2_s((\R^{n-1})^p)$ the subspace of $L^2((\R^{n-1})^p)$ defined by 
\begin{multline*}
L^2_s((\R^{n-1})^p):=\left\{
f^{(p)}\in L^2((\R^{n-1})^p)\ \vert\ \forall (g_1,\ldots,g_p)\in L^2(\R^{n-1});\forall \sigma\in\perm_p\ ;\phantom{\int}
\right.\\ \left.
\int_{(\R^{n-1})^p}f^{(p)}\sur{g_{\sigma(1)}\otimes\cdots \otimes g_{\sigma(p)}}
=\int_{\R^{n-1}}f^{(p)}\sur{g_{\sigma(1)}\otimes\cdots \otimes g_{\sigma(p)}}
\right\}
\end{multline*}
Then we denote by $F_0$ the (algebraic) direct sum 
$$
F_0:={\bigoplus_{p\ge 0}}L^2_s((\R^{n-1})^p)
$$
where $F_0:=\C$. The space $F_0$ has a naturel hermitian structure inherited from $L^2_s((\R^{n-1})^p)$ and 
we denote by $\Fock(L^2_s(\R^{n-1}))$ (or simply $\Fock$) the completion of $F_0$ \ie the set of family 
$(f^{(p)})_{p\in\N}$ of functions $f^{(p)}\in L^2_s((\R^{n-1})^p)$ such that 
$$
\vert f^{(0)}\vert^2\ +\ \sum_{p\ge 1} \int_{(\R^{n-1})^p}\vert f^{(p)}(\vect{x}_1,\ldots,\vect{x}_p)\vert^2 \dd \vect{x}_1
\cdots\dd \vect{x}_p
<+\infty
$$
Hence we get a separable Hilbert space which is calling \emph{symmetric} or \emph{bosonic Fock space}. 

\subsection{Free Quantum Field}\label{Q.libre}
Let $f$ belong to $L^2(\R^{n-1})$ ; we denote by $a^-(f)$ and $a^-(f)^*$ the unbounded operator on $\Fock$ with domain $F_0$ 
such that for all $v=\sum v^{(p)}\in F_0$ where $v^{(p)}\in L^2_s((\R^{n-1})^p)$, we have  
$a^-(f)v:=\sum_p (a^-(f)v)^{(p)}$ and $a^-(f)^*v:=\sum_p(a^-(f)^*v)^{(p)}$ where for all $p\in\N$ 
\begin{align*}
\left(a^-(f)v\right)^{(p)}(m_1,\ldots,m_p):=&\sqrt{p+1}\int_{\R^{n-1}}\sur{f}(m)v^{(p+1)}(m,m_1,\ldots,m_p)\\
\left(a^-(f)^*v\right)^{(p)}(m_1,\ldots,m_p):=&\frac{1}{\sqrt{p}}\sum_{j=1}^p f(m_j)v^{(p-1)}(m^{\vee j})
\end{align*}
where $m^{\vee j}:=(m_1,\ldots,m_{j-1},m_{j+1},\ldots,m_p)$. 
We can remark that for all $f,g\in L^2(\R^{n-1})$, the operators $a^-(f)$ and $a^-(g)^*$ stabilize $F_0$ ; hence we can compose 
them. \\

Then we introduce some notations : $\mu$ denote the function $\mu:\R^{n-1}\to\R^{*+}$ such that 
$$
\mu(\vect{k}):=\left(\vert \vect{k}\vert^2+m^2\right)^{1/2}>0
$$
For all $f\in\mc{S}(\R^{n-1})$ we denote by $Cf\in\mc{S}(\R^{n-1}) $ the function $Cf(\vect{k}):=\sur{f(-\vect{k})}$ and 
$\chapo{f}\in\mc{S}(\R^{n-1})$ denote the Fourier transform of $f$ \ie 
$$
\chapo{f}(\vect{k}):=\frac{1}{(2\pi)^{(n-1)/2}}\int_{\R^{n-1}}e^{-i\vect{k}.\vect{x}}f(\vect{x})\dd\vect{x}
$$\\

Let $f$ belong to $\mc{S}(\R^{n-1})$ ; let $\varphi_m(f)$ and $\pi_m(f)$ denote the unbouded operators on $\Fock$ with domain 
$F_0$ defined by 
\begin{equation}\label{def.phi.m}
\begin{array}{rl}
\varphi_m(f):=&\displaystyle{\frac{1}{\sqrt{2}}
\left[a^-\left(C\chapo{f}/\sqrt{\mu}\right)+a^-\left(\chapo{f}/\sqrt{\mu}\right)^*\right]}\\
\pi_m(f):=&\displaystyle{\frac{i}{\sqrt{2}}
\left[a^-\left(\sqrt{\mu}\chapo{f}\right)^*-a^-\left(\sqrt{\mu}C\chapo{f}\right)\right]}
\end{array}
\end{equation}
One can show (see \cite{Fock.rigoureux.2} p.218) that for all \emph{real valued} $f$ and $g$ the operators 
$\varphi_m(f)$ and $\pi_m(g)$ satisfy the \emph{Canonical Commutation Relation} (CCR)\ie 
\begin{equation}\label{app.CCR}\tag{CCR}
\left[\varphi_m(f),\pi_m(g)\right]=i\left(\int_{\R^{n-1}} fg\right)id
\end{equation}
The operator $\varphi_m(f)$ can be seen as the quantization of the functional $\varphi\mapsto\int_{t=0}f\varphi$ over 
free classical field $\varphi$ and the operator $\pi_m(f)$ as the quantization of 
$\int_{t=0}g\dsurd{\varphi}{t}$. 
\\

Lastly for all $f\in\mc{S}(\R^{n-1})$ we define the operators $\opa(f)$ and $\opa^\dagger(f)$ by
\index{NTO}{Fock.creation@$\opa(f)$}\index{NTO}{Fock.cdestruction@$\opa^\dagger(f)$}
$$
\begin{array}{lcr}
\displaystyle{\opa^\dagger(f):=a^-(\sqrt{\mu}f)^*}&\text{ et }&\displaystyle{\opa(f):=a^-(\sqrt{\mu}Cf)}
\end{array}
$$
Then $\opa(f)$ and $\opa^{\dagger}(g)$ are unbouded operators with domain $F_0$ and one can check that 
\begin{equation}\label{passage}
\begin{array}{rl}
\varphi_m(f):=&\displaystyle{\frac{1}{\sqrt{2}}
\left[\opa(\chapo{f}/\mu)+\opa^\dagger(\chapo{f}/\mu)\right]}\\
\pi_m(f):=&\displaystyle{\frac{i}{\sqrt{2}}
\left[\opa^\dagger(\chapo{f})-\opa(\chapo{f})\right]}
\end{array}
\end{equation}
In other hand (see \cite{Fock.rigoureux.2} p.210) for all \emph{real valued} $f,g\in\mc{S}(\R^{n-1})$ we have 
the following commutation relation 
\begin{equation}\label{app.comm.opa}
\left[\opa(f),\opa^\dagger(g)\right]=\left(\sur{\int_{\R^{n-1}} \mu(\vect{k}){f(-\vect{k})}{g(\vect{k})}\dd\vect{k}}\right)id
\end{equation}

\section{Proof of the proposition \ref{deformation}}\label{dem.1}
First of all it is clear that $1\in P\subset \sym{P}[\hbar]$ is a unit for $\star_\hbar$. 
Let us focus on the associativity of the star product. 

It sufficies to show that $\Psi^k\star_\hbar(\Phi^l\star_\hbar\Xi^m)=(\Psi^k\star_\hbar\Phi^l)\star_\hbar\Xi^m$ 
for all $(k,l,m)\in\N^3$ and $(\Psi^k,\Phi^l,\Xi^m)\in P^{\odot k}\times P^{\odot l}\times P^{\odot m}$. But we have 
by definition 
\begin{equation}\label{base.dem.roudoudou}
\Psi^k\star_\hbar\left(\Phi^l\star_\hbar \Xi^m\right)=\sum_{\alpha\ge
0}\frac{\hbar^\alpha}{\alpha!}\left(\sum_{p+q=\alpha}C^p_\alpha\
\Psi^k M^p \left(\Phi^l M^q \Xi^m\right)\right)
\end{equation} 
where $C^\alpha_\beta$ denote the binomial coefficient $C^\alpha_\beta=\frac{\beta!}{\alpha!(\beta-\alpha)!}$. 

Let $A$ be an integer ; we study the term of order $A$ with respect to $\hbar$ in \ref{base.dem.roudoudou}. 
Hence we consider $(p,q)\in\N^2$ such that $p+q=A$ and we denote by $N$ the integer $N:=k+l+m-2A$. We have 
\begin{multline}\label{un}
\Psi^k M^p\left(\Phi^l M^q \Xi^m\right) = \sum_{\substack{\beta,\alpha\in\lent
0,n-1\rent^p\\\alpha,\beta\in\lent 0,n-1\rent^q}}\left.\int_{\Sigma^p}\right[
\eta^{\beta\alpha}\D_\alpha(z_{1,p})\Psi^k\\
\odot i^p(z_{1,p})
\left.\left(\int_{\Sigma^q}
\eta^{\alpha\beta}\D_\beta(z_{p+1,A})
\Phi^l\odot i^q(z_{p+1,A})\Xi^m\omega_\alpha(z_{p+1,A})\right)\right]\omega_\beta(z_{1,p})
\end{multline}
where for all $\rho\in M^k$ and $1\le a\le b\le k$, $\rho_{a,b}$ denotes $\rho_{a,b}:=(\rho_a,\rho_{a+1}\ldots,\rho_b)\in M^{b-a+1}$. 
Then one can show easily the following combinatorial lemma 
\begin{lemme}\label{lemme.1}\sl{
\begin{enumerate}
\item For all $z,z'\in\Min$ the operators $i(z)$, $i(z')$, $\D_\beta(z)$ and $\D_\beta(z')$ commute. 
\item Let $(a,b)\in\N^2$, $k\in\N$,  $\tilde{\Psi}\in P^{\odot a}$, $\tilde{\Phi}\in P^{\odot b}$ and $z\in \Min^k$ then we have 
\begin{equation}\label{lemme.1.conclusion}
i^k(z)(\tilde{\Psi}\odot
\tilde{\Phi})=\sum_{j=0}^k\sum_{\substack{J\subset\lent 1,
k\rent\\\vert J\vert=j}}i^j(z_J)\tilde{\Psi}\odot
i^{k-j}(z_{\lent 1,k\rent\setminus J})\tilde{\Phi}
\end{equation}
where for all $J=\{a_1,\ldots, a_j\}\subset\lent 1,k\rent$ such that $\vert J\vert =j$ we denote by $z_J$ the $j$-uplet 
$z_J:=(z_{a_1},\ldots,z_{a_j})$. 
\item Identity \refeq{lemme.1.conclusion} is true with $\D_\beta$ instead of $i$. 
\end{enumerate}
}
\end{lemme}
Then using identity \refeq{lemme.1.conclusion} in \refeq{un} one gets 
\begin{multline*}
\Psi^k M^p\left(\Phi^l M^q\Xi^r\right)=
\sum_{j=0}^p\sum_{\beta,\alpha\in\lent 0,n-1\rent^A}
 \sum_{\substack{J\subset\lent 1, p\rent\\\vert
J\vert=j}}
\int_{\Sigma^A} \eta^{\beta\alpha}\D_{\alpha_{1,p}}^p(z_{1,p})\Psi^k\odot \\
\left(i^{p-j}(z_{\lent 1,
p\rent\setminus J})\circ\D^q_{\alpha_{p+1,A}}(z_{p+1,A})\right)\Phi^l\odot
i^{q+j}(z_{J},z_{p+1,A})\Xi^m\omega_\beta(z)
\end{multline*}
and using the Fubini theorem one can reorganize the integration and finally get that the term of order $A$ in 
\refeq{base.dem.roudoudou} is given by 
\begin{multline*}
\sum_{p=0}^A\sum_{j=0}^p\sum_{\beta,\alpha\in\lent 0,n-1\rent^A} C^p_A
C^j_p \int_{\Sigma^A} \eta^{\beta\alpha}\
\D^p_{\alpha_{A-p+1,A}}(z_{A-p+1,A})\Psi^k\odot \\
i^{p-j}(z_{A-p+j+1,A})\circ\D^{A-p}_{\alpha_{1,A-p}}(z_{1,A-p})\Psi^l
\odot i^{A-p+j}(z_{1,A-p+j})\Xi^m\ \omega_\beta(z)
\end{multline*}
Then one can performe the change of variable $j\leftarrow A-p+j$, change the order of summation over $p$ and $j$ and performe 
a second change of variable $p\leftarrow a-A+q$ to get 
\begin{multline*}
\sum_{j=0}^A\sum_{p=0}^j\sum_{\beta,\alpha\in\lent 0,n-1\rent^A}
C^{A-j+p}_A C^{p}_{A-j+p}
\int_{\Sigma^A}\eta^{\beta\alpha}\
\D^{A-j+p}_{\alpha_{j-p+1,A}}(z_{j-p+1,A})\Psi^k\odot \\
i^{A-j}(z_{j+1,A})\circ\D^{j-p}_{\alpha_{1,j-p}}(z_{1,j-p})\Phi^l
\odot i^{j}(z_{1,j})\Xi^m\ \omega_{\beta}(z)
\end{multline*}
But a simple calculation shows that $C^{p+A-j}_A C^{p}_{p+A-j}=C^p_j C^j_A$, hence using the Fubini theorem the last 
expression leads to 
\begin{multline*}
\sum_{j=0}^A C^j_A\sum_{p=0}^j\sum_{\beta,\alpha\in\lent 0,n-1\rent^A}
\sum_{\substack{J\subset\lent 1\ldots j\rent\\\vert J\vert=p}}\int_{\Sigma^A}\eta^{\beta\alpha}
\D^p_{\alpha_J}(z_J)\left[\D^{A-j}_{\alpha_{j+1,A}}(z_{j+1,A})\Psi^k\right]\odot\\
\D^{j-p}_{\alpha_{\lent 1, j\rent\setminus J}}(z_{\lent 1,
j\rent\setminus J}) \left[i^{A-j}(z_{j+1,A})\Phi^l\right]
\odot i^{j}(z_{1,j})\Psi^m\omega_{\beta}(z)
\end{multline*}
Thus lemma \ref{lemme.1} gives the final result. 

\section{Operator Ordering}\label{dem.2}
We will prove the theorem \ref{ordre}. First of all, using the equal times commutation relation \refeq{CCR} and 
following the proof of the well known \emph{Wick lemma} (see \eg \cite{RYDER}), one can prove the following 
\begin{lemme}\label{Wick.appendice}\sl{
Let $(a,b,c,d)\in\N^4$ and $f^{(a)}\in\mc{S}((\R^{n-1}))^a$, $f^{(b)}\in\mc{S}((\R^{n-1}))^b$, $f^{(c)}\in\mc{S}((\R^{n-1}))^c$ 
and $f^{(d)}\in\mc{S}((\R^{n-1}))^d$ be real valued functions ; for all $p\in\N^*$ and all $f^{(p)}\in\mc{S}(\R^{n-1})^p$, 
we define $\pi_m^{\otimes p}(f^{(p)}):=\prod_{j=1}^p\pi_m(f^{(p)}_j)$ and 
$\varphi_m^{\otimes p}(f^{(p)}):=\prod_{j=1}^p\varphi_m(f^{(p)}_j)$. Then $\pi_m^{\otimes p}(f^{(p)})$ and 
$\varphi_m^{\otimes p}(f^{(p)})$ are unbounded operator acting on $\Fock$ with domain $F_0$ which stabilize this domain. 

Then the operator production $\pi_m^{\otimes a}(f^{(a)})\varphi_m^{\otimes b}(f^{(b)})\cdot 
\pi_m^{\otimes c}(f^{(c)})\varphi_m^{\otimes d}(f^{(d)})$ is given by the following expression 
\begin{multline}\label{expr.Wick}
\sum_{I\subset \lent 1,b\rent}\sum_{\substack{\sigma:I\hookrightarrow\lent 1,c\rent\\ \sigma\text{ injective}}}i^{\vert I\vert}
\left(\prod_{j\in I}\int_{\R^{n-1}}f^{(b)}_j f^{(c)}_{\sigma(j)}\right)\\
\pi_m^{\otimes(a+c-\vert I\vert)}\left(f^{(a)},f^{(c)}_{\lent 1,c\rent\setminus\sigma(I)}\right)
\varphi_m^{\otimes(d+b-\vert I\vert)}\left(f^{(d)},f^{(b)}_{\lent 1,b\rent\setminus I}\right)
\end{multline}
where for $p=c$ or $d$ and for all $J\subset \lent 1,p\rent$ we write 
$f^{(p)}_{J}$ for $(f^{(c)}_j)_{j\in J}\in \mc{S}((\R^{n-1}))^{\vert J\vert}$. 
}
\end{lemme}

To prove the theorem \ref{ordre} we have to show the following identity 
\begin{equation}\label{ce.quon.cherche.ordre}
\Theta(\Psi^k)\Theta(\Phi^l)=\Theta(\Psi^k\star_\hbar \Phi^l)
\end{equation}
for all $(k,l)\in\N^2$, $\Psi^k\in P^{\odot k}$ and all decomposable element $\Phi^l\in P^{\odot l}$.

\begin{proof}(of theorem \ref{ordre}). 
Let $(k,l)\in\N^2$, $\Psi^k=\psi_1\odot\cdots\odot\psi_k\in P^{\odot k}$ and 
$\Phi^l=\phi_1\odot \cdots\odot\phi_l\in P^{\odot l}$. Then we have 
\begin{multline}\label{base.dem.ordre}
\Theta(\Psi^k)\Theta(\Phi^l)=\\
\sum_{I\subset\lent 1,k\rent}\sum_{J\subset\lent 1,l\rent}(-1)^{k-\vert I\vert+\vert J\vert}
\pi_m\left(\psi_{\lent 1,k\rent\setminus I}\right)\varphi_m\left(\dsurd{\psi_I}{t}\right)
\pi_m\left(\phi_J\right)\varphi_m\left(\dsurd{\phi_{\lent 1,l\rent\setminus J}}{t}\right)
\end{multline}
where for all $v=v_1\odot\cdots\odot v_q\in\PP^{\odot q}$ and all $I\subset\lent 1,q\rent$ we denote by $v_I$ the element of 
$\PP^{\odot \vert I\vert}$ defined by $v_I:=\bigodot_{i\in I}v_i$ and where for all 
$w=w_1\odot\cdots\odot w_q\in\PP^{\odot q}$ we set $\pi_m(w):=\prod_{i=1}^q\pi_m((w_i)_{\vert t=0})$ and 
$\varphi_m\left(\dsurd{w}{t}\right):=\prod_{i=1}^q\pi_m\left(\dsurd{w_i}{t}_{\vert t=0}\right)$. 

Then using lemma \ref{Wick.appendice} the right hand side of \refeq{base.dem.ordre} is given by 
\begin{multline*}
\sum_{I\subset\lent 1,k\rent}\sum_{J\subset\lent 1,l\rent}\sum_{K\subset I}\sum_{\sigma:I\hookrightarrow J}
i^{\vert K\vert}(-1)^{k-\vert I\vert+\vert J\vert}
\left[\prod_{j\in K}\int_{\R^{n-1}}\dsurd{\psi_j}{t}\phi_{\sigma(j)}\right]\\
\pi_m\left(\psi_{\lent 1,k\rent\setminus I}\odot\phi_{J\setminus\sigma(K)}\right)
\varphi_m\left(\dsurd{\psi_{I\setminus K}}{t}\odot\dsurd{\phi_{\lent 1,l\rent\setminus J}}{t}\right)
\end{multline*}
One can change the order of the sum over  $K,\sigma$ and $I,J$ in order to get 
\begin{multline*}
\sum_{K\subset\lent 1,k\rent}\sum_{\sigma:K\hookrightarrow\lent 1,l\rent}(-i)^{\vert K\vert}
\left[\prod_{j\in K}\int_{\R^{n-1}}\dsurd{\psi_j}{t}\phi_{\sigma(j)}\right]\\
\sum_{\substack{I\subset\lent 1,k\rent\setminus K\\ J\subset\lent 1,l\rent\setminus \sigma(K)}}
(-1)^{\vert I\vert+l-\vert J\vert-\vert K\vert}
\pi_m\left(\psi_I\odot\phi_{(\lent 1,l\rent\setminus \sigma(K))\setminus J}\right)
\varphi_m\left(\dsurd{\psi_{(\lent 1,k\rent\setminus K)\setminus I}}{t}\odot \dsurd{\phi_J}{t}\right)
\end{multline*}
where we recognize $\Theta\left(\psi_{\lent 1,k\rent\setminus I}\odot \phi_{\lent 1,l\rent\setminus \sigma(K)}\right)$. 
Hence we finally get the following identity 
\begin{equation}\label{Theta.ordre}
\Theta(\Psi^k)\Theta(\Phi^l)=
\sum_{K\subset\lent 1,k\rent}\sum_{\sigma:K\hookrightarrow\lent 1,l\rent}(-i)^{\vert K\vert}
\left[\prod_{j\in K}\int_{\R^{n-1}}\dsurd{\psi_j}{t}\phi_{\sigma(j)}\right]
\Theta\left(\psi_{\lent 1,k\rent\setminus K}\odot \phi_{\lent 1,l\rent\setminus \sigma(K)}\right)
\end{equation}

If one writes explicitely the expression of $\Psi^k\star_\hbar\Phi^l$ using the definition of $\D^p(z)$ and $i^p(z)$ 
then one gets 
\begin{equation}\label{fin.comparaison}
\Psi^k\star_\hbar\Phi^l=
\sum_{p\ge 0}\hbar^p\sum_{\substack{J\subset \lent 1,k\rent\\\vert J\vert =p}}
\sum_{\sigma:J\hookrightarrow\lent 1,l\rent}
\left[\prod_{j\in J}\int_{\R^{n-1}}\dsurd{\psi_j}{t}\phi_{\sigma(j)}\right]
\psi_{\lent 1,k\rent\setminus J}\odot \phi_{\lent 1,l\rent\setminus \sigma(J)}
\end{equation}
Then \refeq{fin.comparaison} together with \refeq{Theta.ordre} leads to \refeq{ce.quon.cherche.ordre}. 
\end{proof}

\bibliographystyle{acm}

\end{document}